\documentclass[fleqn,usenatbib]{mnras}

\usepackage{newtxtext,newtxmath}
\usepackage[T1]{fontenc}

\DeclareRobustCommand{\VAN}[3]{#2}
\let\VANthebibliography\thebibliography
\def\thebibliography{\DeclareRobustCommand{\VAN}[3]{##3}\VANthebibliography}
\defcitealias{Scaramella}{EC22}

\usepackage{graphicx}	% Including figure files
\usepackage{amsmath}	% Advanced maths commands
\usepackage{lineno}

\title{Type Ia Supernova cosmology combining data from the \textit{Euclid} mission and the Vera C. Rubin Observatory}

\author[A. Bailey, et al.]{
A. Bailey,$^{1}$\thanks{E-mail: ava.bailey@duke.edu}
M. Vincenzi,$^{1}$\thanks{E-mail: maria.vincenzi@duke.edu}
D. Scolnic$^{1}$,
J.-C. Cuillandre$^{2}$,
J. Rhodes$^{3}$,
E. R. Peterson$^{1}$
B. Popovic$^{1}$
\\
$^{1}$Department of Physics, Duke University Durham, NC 27708, USA\\
$^{2}$Université Paris-Saclay, Université Paris Cité, CEA, CNRS, AIM, 91191, Gif-sur-Yvette, France \\
$^{3}$Jet Propulsion Laboratory, California Institute of Technology
}
\def\snana{\texttt{SNANA}}
\def\euclid{\textit{Euclid}}

\date{Submitted to MNRAS}

\pubyear{2022}

\begin{document}
\label{firstpage}
\pagerange{\pageref{firstpage}--\pageref{lastpage}}
\maketitle

\begin{abstract}
The \euclid\ mission will provide first-of-its-kind coverage in the near-infrared over deep (three fields, $\sim$10--20 square degrees each) and wide ($\sim$10000 square degrees) fields. While the survey is not designed to discover transients, the deep fields will have repeated observations over a two-week span, followed by a gap of roughly six months. In this analysis, we explore how useful the deep field observations will be for measuring properties of Type Ia supernovae (SNe Ia).
Using simulations that include \textit{Euclid}'s planned depth, area and cadence in the deep fields, we calculate that more than 3700 SNe between $0.0<z<1.5$ will have at least five \euclid\ detections around peak with signal-to-noise ratio larger than 3. 
While on their own, \euclid\ light curves are not good enough to directly constrain distances, when combined with LSST deep field observations, we find that uncertainties on SN distances are reduced by 20--30\% for $z<0.8$ and by 40--50\% for $z>0.8$.

Furthermore, we predict how well additional \euclid\ mock data can be used to constrain a key systematic in SN Ia studies --- the size of the luminosity `step' found between SNe hosted in high mass ($>10^{10} M_{\odot}$) and low mass ($>10^{10} M_{\odot}$) galaxies. This measurement has unique information in the rest-frame NIR. We predict that if the step is caused by dust, we will be able to measure its reduction in the NIR compared to optical at the 4$\sigma$ level. 

We highlight that the LSST and \euclid\ observing strategies used in this work are still provisional and some level of joint processing is required.  Still, these first results are promising, and assuming \euclid\ begins observations well before the Nancy Roman Space Telescope (Roman), we expect this dataset to be extremely helpful for preparation for Roman itself.
\end{abstract}


%%%%%%%%%%%%%%%%%%%%%%%%%%%%%%%%%%%%%%%%%%%%%%%%%%

%%%%%%%%%%%%%%%%% BODY OF PAPER %%%%%%%%%%%%%%%%%%

\section{Introduction}

 The \euclid\ survey is a space-based mission with the primary goal of understanding the accelerating expansion of the universe and the nature of its components.  Its primary science is optimized for two complementary cosmological probes: weak gravitational lensing and baryonic acoustic oscillations.  The telescope will allow for both high-precision photometric imaging in the near-infrared (NIR) as well as spectroscopy.  While \citet{Astier14} proposed for a transient survey for \euclid\ that would enable a Stage IV dark energy measurement \citep{Albrecht06} with Type Ia supernovae (SNe Ia), this survey is currently unplanned. In this paper, we study what may still be leveraged in terms of discovery and measurements of SNe Ia, given a survey that is not designed for such a study but is still unique in its capabilities. 

The majority of surveys used for SN Ia cosmological studies are conducted in the optical wavelength range \citep[ex., Pantheon+;][]{Scolnic21,Brout22}. In the NIR, there have been a handful of surveys that discover or follow-up SNe Ia at low redshift ($z<0.08$) like CSP \citep{Hamuy06}, CfA \citep{Wood-Vasey08,Friedman15}, DEHVILS (Peterson et al.~in prep.), Do et al.~in prep., but only the RAISIN survey \citep{Jones22} measured light curves of SNe at intermediate $z$ ($0.3<z<0.5$). The RAISIN program combined the NIR data with light curves measured in the optical from Pan-STARRS \citep{Chambers16} and the Dark Energy Survey \citep{DES_spec}, and the total number of light curves usable for cosmological studies was 37. Therefore, even a small number of light curves at higher redshift measured in the NIR would be a significant contribution to the field (ex., SUSHI; \textit{HST}-GO 15363, PI: Suzuki).

The challenge for a transient survey with \euclid\ is its cadence.
Typically, surveys of SNe Ia have cadences of 5--10 days in order to measure the rise, peak and decline of the light curve in multiple passbands.  In \textit{Euclid}'s deep fields, as discussed below, there is a series of observations over approximately a week and then a gap of six months.  The only works so far that have studied the viability of a transient survey with poor cadence optimization for SN studies are \citet{Inserra18}, \citet{2022arXiv220408727M} and \citet{2022arXiv220409402T}, which are focused on long-duration transients like pair-instability SNe and superluminous SNe. We build off these works but change the focus to SNe Ia and explore the possibility of combining \euclid\ with the ground-based Legacy Survey of Space and Time \citep[LSST;][]{2019ApJ...873..111I}.
LSST, conducted on the Vera C. Rubin Observatory (hereafter Rubin), will measure high quality optical light curves for $>1$ million SNe Ia. It is expected to have significant overlap in time and footprint with the \euclid\ mission \citep{Capak19}, and this could yield a large SN sample with high quality optical and NIR data.

To enable a study of the usefulness of \euclid\ measurements of SNe Ia, we use the SuperNova ANAlysis software package \citep[\snana; ][]{Kessler09}, which has been used for a large number of survey forecasts \citep{Jones_2017_I, Kessler19, DES_biascor, Vincenzi21, 2021ApJ...913...49P}. We also rely on new work that extends the SN Ia spectral model into the NIR from recent studies using their SALT3 framework \citep{Pierel18,Kenworthy21,Pierel22}.
For \euclid\ simulations, we use the \euclid\ observing strategy and survey description presented by \citet{Scaramella} (hereafter \citetalias{Scaramella}). 
For LSST simulations, we leverage recent work in \citet{LSST21} and \citet{Sanchez22} and recreate this work at the catalog level.\color{black} 
While the number of SNe measured with \euclid\ is likely a small fraction of the many thousands of SNe measured by LSST, the NIR information from \euclid\ can provide key tests of systematics that will enhance the dark energy constraints from LSST.  One such area, in particular, is a better understanding of the intrinsic scatter of SNe Ia, which explains residual dispersion in standardized brightnesses after accounting for measurement uncertainties.  One of the leading explanations for this scatter is that it is due to variations in the reddening ratios in different galaxies \citep{BS20}. A prediction from the \citet{BS20} model is that in the NIR, correlations seen between standardized brightnesses and host galaxy properties should disappear \citep{Uddin20,Johansson21,Ponder21}.  As samples used to evaluate this prediction with NIR data are small and mostly at low-$z$ with specific selection biases, \euclid\ could make a critical measurement for future SNe Ia analyses.

Furthermore, SN data from \euclid\ can be an excellent preparatory sample for the Nancy Grace Roman Survey Telescope \citep[Roman; ][]{2018ApJ...867...23H, 2021arXiv211103081R}, which has been specifically designed for SNe Ia observations.  For Roman, the cadence is $\sim$5 days, which will make it ideal for discovering and measuring SNe Ia. Still, as \euclid\ will launch more than a couple years before Roman, this could be a fantastic opportunity for Roman preparation.  These statements all rely on whether \euclid\ data will be processed to discover and measure SNe.  One of the main goals of this analysis is to advocate for this ability. 
The structure of this paper is as follows.  In Section~\ref{sec:simulations}, we describe the \euclid\ and LSST surveys and the catalog-level simulations.  In Section~\ref{sec:SNquality}, we show expected light curves as observed with \euclid\ along with numbers of SNe measured and distance constraints.  In Section~\ref{sec:Mass_Step}, we forecast constraints on correlations between supernova properties and host-galaxy properties using \textit{Euclid}. Our final remarks are in Section~\ref{sec:Discussion}.

\section{Simulations}\label{sec:simulations}

\subsection{Simulation Framework}

\label{sec:sim_general}

\begin{figure*}
    \centering
    \includegraphics[width=\textwidth]{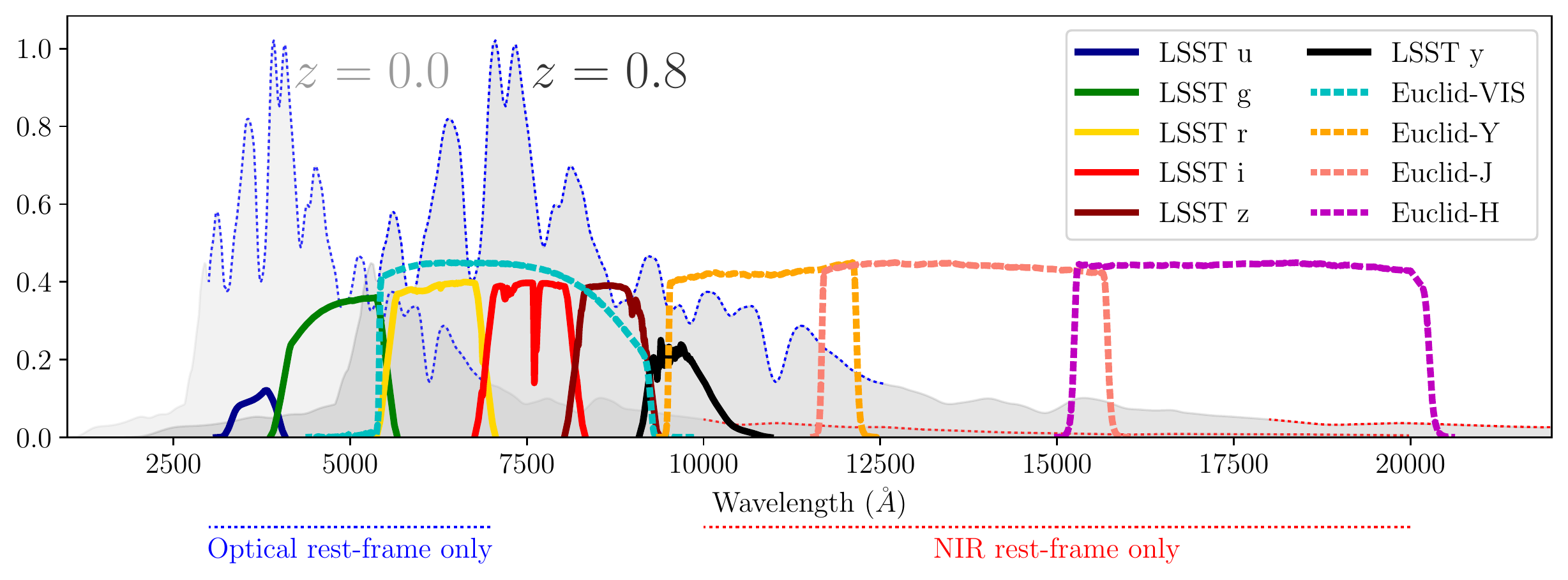}
    \caption{Transmission function of the LSST filters ($ugrizy$) and the \euclid\ filters (VIS+$YJH$). We compare the spectral energy distribution at peak brightness of a SN Ia at $z=0$ and $z=0.8$.}
    \label{fig:bands_specia}
\end{figure*}

\begin{figure*}
    \centering
    \includegraphics[width=\textwidth]{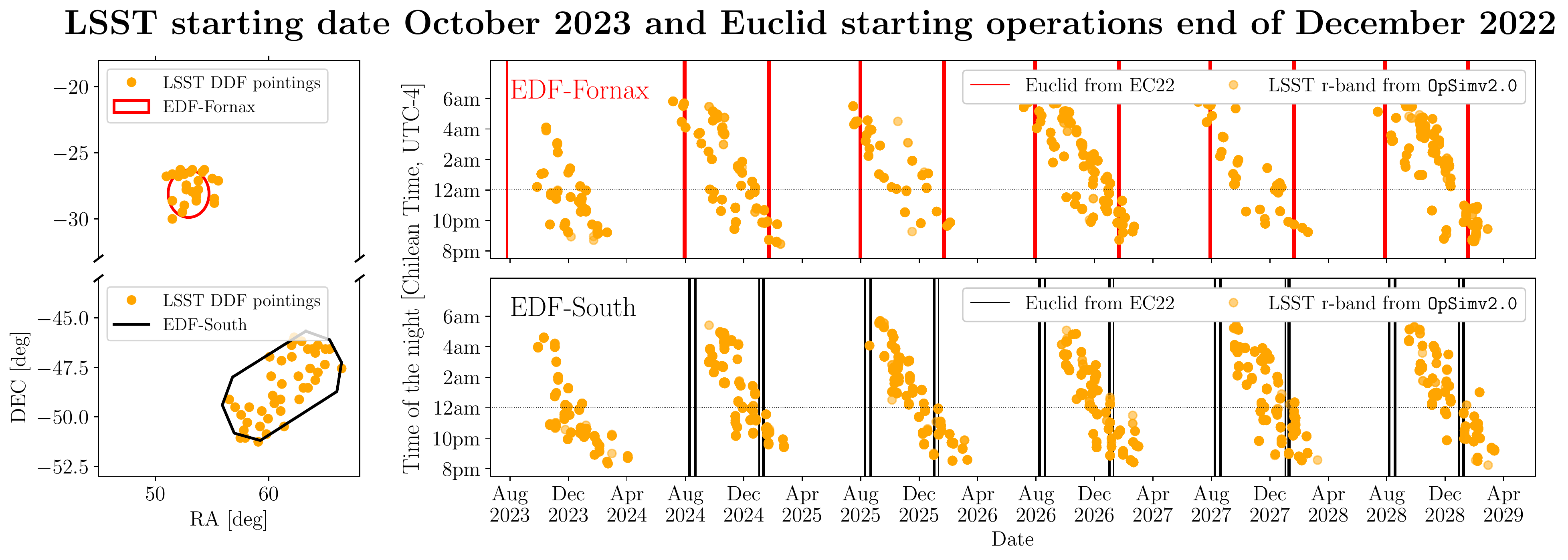}
    \caption{Comparison between \euclid\ and LSST mock observing strategies in the Deep Fornax and Deep South fields. \textit{Left:} Footprint of the \euclid\ Deep Fornax (red) and Deep South fields (black) footprints, compared to LSST pointings in the equivalent deep drilling fields. \textit{Right:} Distribution of LSST $r$-band observations as a function of time of the night (in Chilean time, i.e., UTC-4) over five years and distribution of \euclid\ observations (red and black vertical lines, note that Euclid observes over 24 hours). LSST observing strategy is from \texttt{OpSim-v2.0} and \euclid\ observing strategy is the `Euclid reference survey' from \citetalias{Scaramella}. EDF-Fornax field will be observed by \euclid\ every six months for 6 consecutive nights, EDF-South will be observed by \euclid\ every 5 and 7 months, 3--4 times over a time range of 13/14 days. The EDF-Fornax and EDF-South will be visible from the Rubin Observatory only between August and February.}
    \label{fig:cadence}
\end{figure*}

Simulations used in this analysis are generated and analysed using the \snana\ software package \citep{Kessler09}.\footnote{\url{https://github.com/RickKessler/SNANA}.} \snana\ is an open-source package designed to generate catalogue level simulations of transient surveys. In this analysis, we only simulate SNe Ia and their host galaxies, and we assume a flat $\Lambda$CDM cosmological model with Hubble constant $H_0$ = 70 km s$^{-1}$ Mpc$^{-1}$ and $\Omega_M= 0.311$.

The main steps that constitute the \snana\ simulation are the following. First, SN Ia volumetric rate measurements are used to estimate the absolute number of SNe expected to explode within a volume and time interval. In this work, we model rates following \citet{2018ApJ...867...23H} and consider a time window of five years (out of the 6-year duration of the Euclid mission, when the two southern deep fields will be regularly observed) and a redshift range of $0<z<1.5$. SNe Ia are then simulated using the SN Ia SALT3 model presented by \citet{Kenworthy21} and extended into the NIR by \citet{Pierel22}. This model defines the spectral energy distribution (SED) of a SN Ia within a wavelength range of $2000$--$20000$ \AA\ and between a phase range of $-$15 to $+$45 (rest frame) days from peak brightness. 

The SALT3 model is described by five parameters: SN redshift $z$, an amplitude term $x_0$, SN light-curve `stretch' term ($x_1$), SN restframe $B-V$ color at peak ($c$) and time of peak brightness $t_0$. In our simulations, SN peak times are uniformly generated across the six years of the mission, and SN $c$ distributions are simulated following \citet{Scolnic_2016}. We generate SNe Ia brightnesses ($x_0$) so that their intrinsic scatter in luminosity follows \citet{Guy10} and we parametrize the linear luminosity-stretch correlation ($\alpha$) and luminosity-color correlations ($\beta$) assuming slopes of $\alpha=0.14$ and $\beta=3.1$.

After generating the sample of SNe Ia and associated SED models, various astrophysical effects are applied, including redshifting, cosmological dimming and dust extinction from host galaxy and Milky Way. We use a \cite{Cardelli89} dust law with $R_V=3.1$ for Milky Way dust extinction.

Each simulated SN is then associated to a host galaxy, which in this analysis is chosen using the galaxy catalogue presented by \citet{DES_massstep}. This catalogue was generated from DES Science Verification data and includes $\sim$380,000 galaxies at $0<z<1.5$. Galaxy association is implemented following measurements of SN Ia rates as a function of galaxy properties \citep[as described in ][]{Vincenzi21}. We use the rates presented by \citet{Wiseman_rates} and simulate correlations between SN stretch and SN host galaxy stellar mass following \citet{2021ApJ...913...49P}.

Given the final SED model and SN host galaxy, the \lq true\rq\ broadband photometry of each simulated source is estimated by integrating the SN SED over the survey filters (in this analysis, the LSST and \euclid\ filters illustrated in Fig.~\ref{fig:bands_specia}). From the \lq true\rq\ SN photometry, we estimate the \lq observed\rq\ SN photometry by applying observational noise and adding background light from the host galaxy. This step uses the observing conditions provided in a pre-computed observational library (referred to as a \lq \texttt{SIMLIB}\rq) that includes information on cadence, zeropoints, sky noise and PSF size for every night of observation, for every survey. In Sec.~\ref{sec:sim_euclid} and \ref{sec:sim_lsst}, we discuss the LSST and \euclid\ observing libraries used in our simulations.

\subsection{Simulations of the \euclid\ Deep Survey}
\label{sec:sim_euclid}

We simulate SN Ia light curves as observed by the \euclid\ satellite according to planned observations of its Deep Field survey.
 The \euclid\ telescope will be equipped with a visible imager, with one broad filter referred to as \textit{VIS}, a near-infrared imager with \textit{Y, J, H} filters \citep[see Fig.~\ref{fig:bands_specia} and ][]{2022A&A...662A..92E} and NIR grisms. \euclid\ will perform  a `wide' survey (15000 sq. deg.) and a `deep' survey (50 sq. deg.) over six years. 

There are three main deep fields planned: \euclid\ Deep Survey North (EDF-North, 20 sq. deg.), \euclid\ Deep Survey South (EDF-South, 23 sq. deg.) and \euclid\ Deep Survey Fornax (EDF-Fornax, 10 sq. deg.). Every six months, EDF-South and EDF-Fornax will be repeatedly observed for two weeks with a relatively homogeneous cadence of 2 to 4 days, whereas EDS-North is scheduled for inhomogeneous cadence with consecutive visits ranging from 16 to 55 days. Therefore, for this analysis, we only consider EDF-South and EDF-Fornax. EDF-Fornax will be observed every six months for 6 consecutive nights, while EDF-South will be observed for 3--4 times over a time range of 13 days, every six months \citep[see][for more details]{Inserra18}.

For \euclid, we use the `\euclid\ reference survey definition' presented by \citetalias{Scaramella} (see fig.~29 and sec.~8). This is the latest version of the full \euclid\ schedule, and it includes observing cadence for both the wide and deep \euclid\ surveys, as well as observations of calibration fields. This is the most up-to-date, publicly available schedule for \euclid. It predicts that \euclid\ will observe EDF-South and EDF-Fornax in August and February every year, i.e., at the beginning and towards the end of each LSST season (see Fig.~\ref{fig:cadence} and Sec.~\ref{sec:sim_lsst}). In Sec.~\ref{sec:cadence_uncertaint.}, we discuss the caveats related to potential changes in the \euclid\ reference survey.

The two \euclid\ Deep Fields, EDS-South and EDS-Fornax, have excellent overlap with the LSST Deep Drilling Fields (DDFs, see Fig.~\ref{fig:cadence}) and similar depth (5$\sigma$ limiting magnitude of 26 for $VIS$ and 24.1--24.5 for $YJH$, see \citetalias{Scaramella}). Thus, there is potential for synergies with LSST transient science, despite the \euclid\ observing strategy not being optimized for it.

\subsection{Simulations of the LSST Deep Drilling Fields}
\label{sec:sim_lsst}
Following the same approach and software used for \euclid, we simulate SN Ia light curves as observed by LSST. LSST is an optical imaging survey conducted on the Vera Rubin Observatory. It includes a Wide Fast Deep survey (WFD, 18,000 sq. deg.) and 5 Deep Drilling Fields (DDF, 100 sq. deg. total). In this analysis, we focus on the two LSST DDFs Fornax and South, that overlap with \euclid. 
The overlap between the two surveys in these fields spans 10 sq. deg. and 23 sq. deg., respectively \citep{Capak19}. We note that the LSST DDF South was not originally planned and has been recently added due to the strong synergistic science possibilities presented by \citet{2022zndo...5836022G}.\footnote{The `Rubin-Euclid Derived Data Products: Initial Recommendations' document. See also announcement in \url{https://community.lsst.org/t/scoc-endorsement-of-euclid-deep-field-south-observations/6406}.}
  
The cadence, exposure times and observational noise for the considered LSST DDFs are simulated based on the output of the LSST Operations Simulation software. This software computes the LSST pointing history and relative observing conditions based on the LSST Feature-Based Scheduler.
In our analysis, we use the LSST Operations Simulation version 2.0 (baseline, \texttt{OpSim-v2.0}),\footnote{The simulations can be downloaded at \url{https://epyc.astro.washington.edu/~lynnej/opsim_downloads/fbs_2.0/baseline/}, \texttt{OpSim-v2.0} Summary Information Document and release can be found at \url{https://community.lsst.org/t/survey-simulations-v2-0-release-nov-2021/6059} and links therein. References for the LSST operation simulations: \citet{2014SPIE.9150E..15D}, \citet{2016SPIE.9910E..13D}, \citet{2016SPIE.9911E..25R}.} and translate it into \snana-readable format using the package published by \citet{2020ApJS..247...60B}.  We note that this cadence is not yet set, and further optimization of the observing strategy is ongoing \citep{Gris22}.

For this mock observing strategy, the average cadence in the LSST DDFs is 3--4 days in $ri$ and 5 days in $ugzy$, and the 5$\sigma$ limiting magnitude is approximately 26th magnitude. In Fig.~\ref{fig:cadence}, we compare the distribution of LSST observations in the Fornax and South deep fields (all filters combined) with the distribution of observations from \euclid.

\begin{figure}
    \centering
    \includegraphics[angle=90,width=1.05\columnwidth]{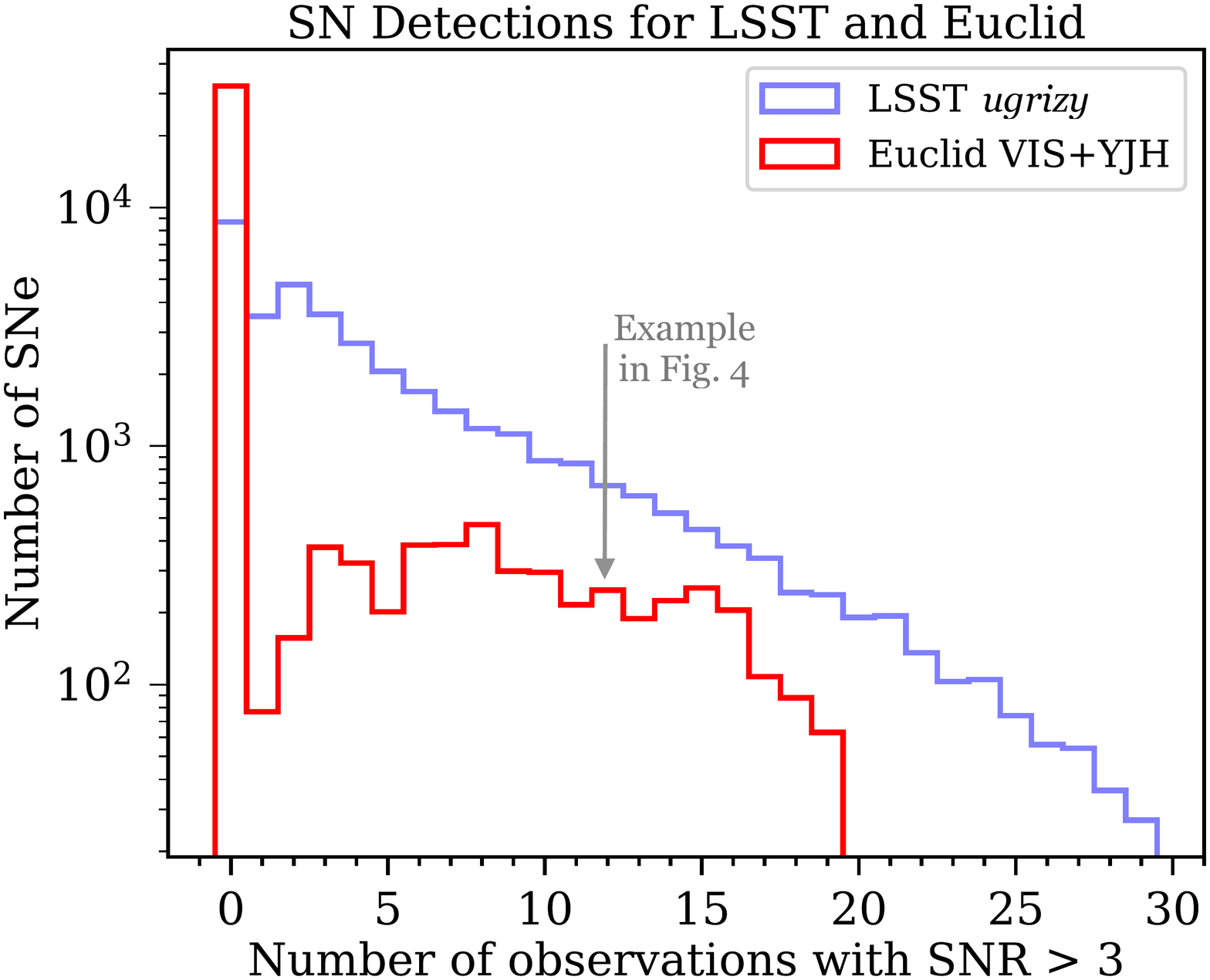}\\
    \caption{Number of data points with signal-to-noise ratio (SNR) larger than 3 in any LSST filter (blue empty histogram) and in any \euclid\ filter (empty red histogram). We consider only the (rest-frame) phase window of $-15$ to +50 days. No SN will have more than 20 observations from \euclid. This is because Euclid observes for a maximum of five nights (in 4 filters) every six months. In our simulations, we only record SNe with at least one data point with SNR>3 in LSST \textit{griz} filters; all other SNe are ignored.}
    \label{fig:detections}
\end{figure}

\begin{figure}
    \centering
    \includegraphics[width=0.9\columnwidth]{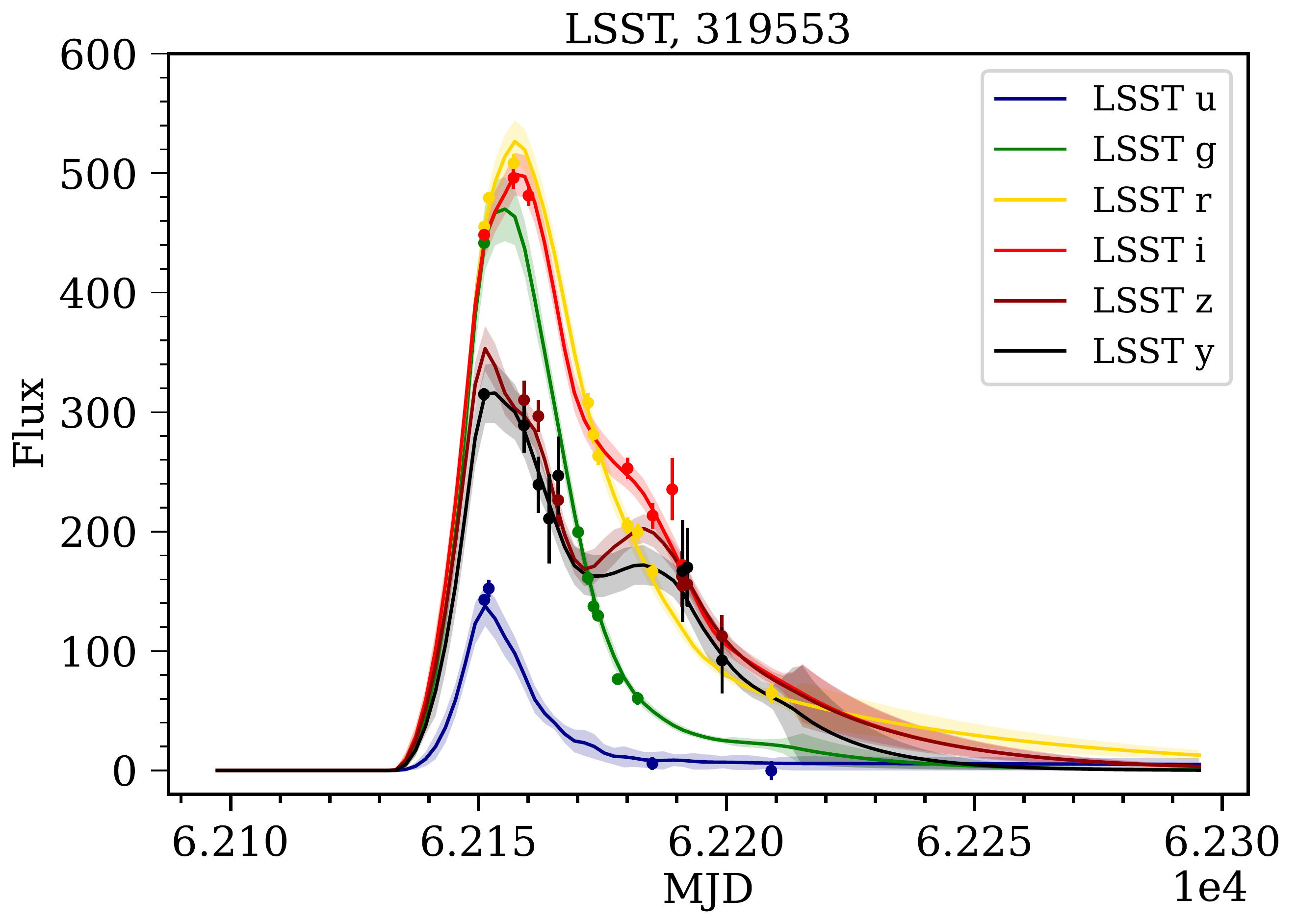}\\
    \includegraphics[width=0.9\columnwidth]{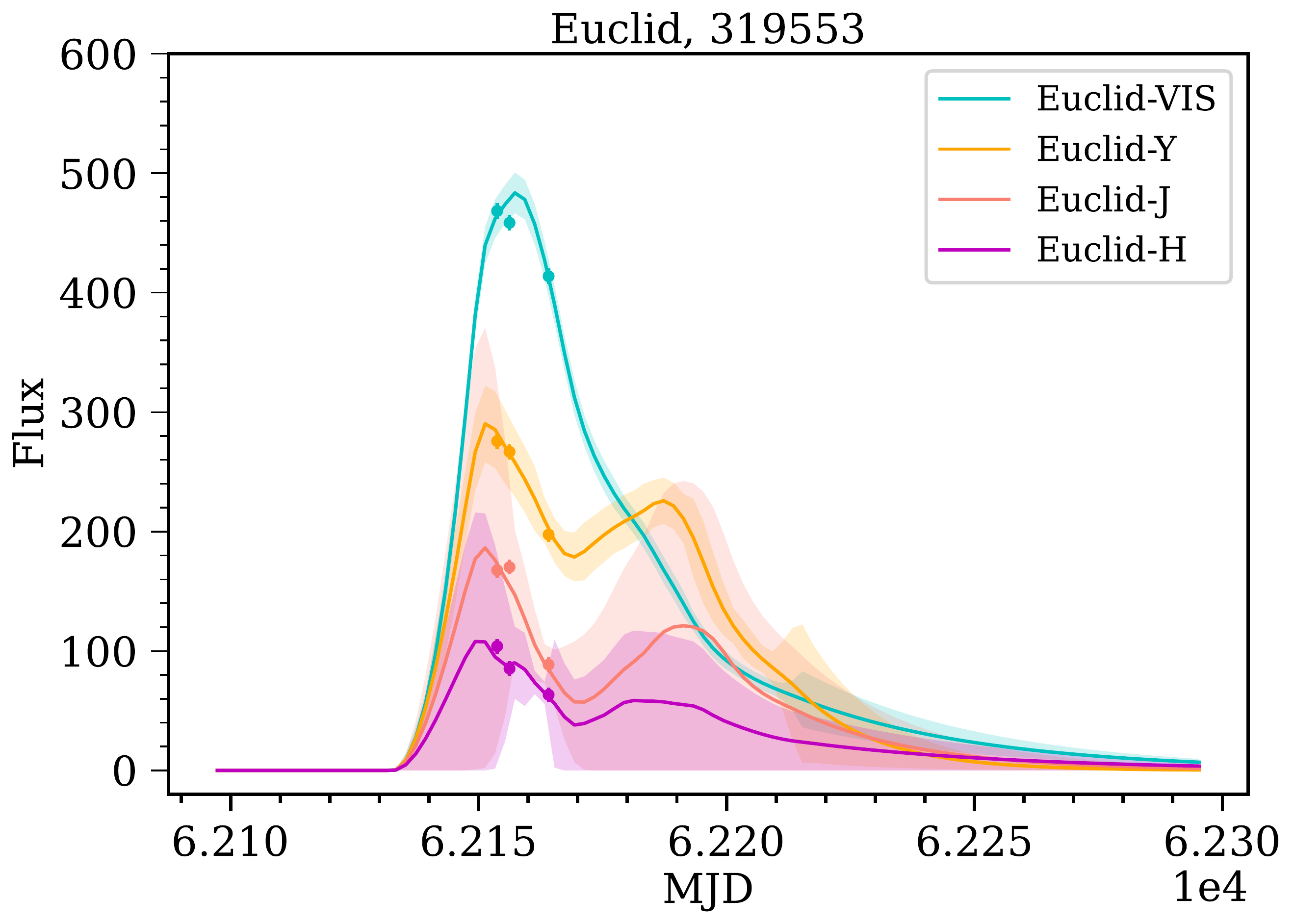}\\
    \caption{Example of a SN Ia (ID: 319553) light curve simulated as seen with LSST (\emph{top}) and as seen with \euclid\ (\emph{bottom}). We provide these simulated light curves as examples of what output can be expected from plotting real data light curves.}
   \label{fig:LC}
\end{figure}

\subsection{Uncertainties on the LSST and \euclid\ observing strategies}
\label{sec:cadence_uncertaint.}

Our analysis uses publicly available observing strategies both for \euclid\ (\citetalias{Scaramella}) and for LSST \citep{2014SPIE.9150E..15D, 2016SPIE.9910E..13D, 2016SPIE.9911E..25R}. The `Euclid reference survey' from \citetalias{Scaramella} assumes that \euclid\ will launch in October 2022 and begin survey operations in December 2022. The LSST \texttt{OpSim-v2.0} assumes that the beginning of survey operations for Rubin will be in October 2023.

Both the LSST and \euclid\ surveys have been subject to significant delays in the last couple of years, therefore these publicly available starting dates are outdated. For LSST, the starting date is likely to be delayed by approximately one year (beginning of the survey in October 2024). For \euclid, the starting date is still uncertain and the launch is currently being rescheduled for 2023.

Despite the uncertain starting date and observing strategies for both LSST and \euclid, we do not expect the results of our analysis to change. The visibility windows for the two fields considered in this analysis are very well defined, both for Rubin and for \euclid. Therefore, we expect the overlap between the two surveys presented in Fig.~\ref{fig:cadence} to be unaffected by shifts in the surveys' starting dates. We further justify this statement below.

The visibility interval of the EDF-South and EDF-Fornax from Rubin is between August and February, independently on which year operations will start. This is shown in Fig.~\ref{fig:cadence}. In August, the two deep fields start to become visible towards the end of the night. They remain observable until February, when they can be observed only at the very beginning of the night.

\euclid\ restrictions on the visibility windows for EDF-South and EDF-Fornax  come from the fact that the telescope always must observe at 90 degrees from the Sun (forward and backwards, see sec. 6.3. in \citetalias{Scaramella}). For this reason, we can confidently assume that the time of year during which \euclid\ will observe EDF-South and EDF-Fornax will be roughly August and January. Even though generating an updated Euclid schedule would require the (not yet publicly available) scheduling software \texttt{ECTile} (see sec. 6.3. and 7 in \citetalias{Scaramella}), we expect minor changes compared to the `\euclid\ reference survey' schedule (in particular, for the EDF-Fornax and EDF-South observations).

\subsection{Auxiliary spectroscopic observations}
In our analysis, we assume spectroscopic redshifts are available for all measured SNe Ia. We expect most of these spectroscopic redshifts to be observed through the ground-based Time-Domain Extragalactic Survey \citep[TiDES, ][]{2019Msngr.175...58S} on the multi-object spectrograph 4MOST \citep{2019Msngr.175....3D}. TiDES is expected to perform comprehensive spectroscopic follow-up of LSST SN host galaxies both in the wide and deep LSST fields. The \euclid\ telescope will also measure spectroscopic redshifts using NIR grisms \citep{2022A&A...662A..92E}. However, \euclid\ spectroscopic redshifts will only be available at the end of the survey.

\color{black}
\section{SN light-curve quality}
\label{sec:SNquality}
In this section, we present the number and quality of SN Ia light curves simulated for LSST and \euclid. When running our simulations, we only record SNe that have at least one detection with signal-to-noise ratio (SNR) larger than 3 in one of the LSST filters at any phase.

\subsection{SN Detection} \label{sec:SN_Detect}
For each simulated SN, we consider a phase window of $-$15 to +50 (rest-frame) days from peak and we estimate the number of observations with SNR larger than 3 from \euclid\ (all filters combined) and from LSST (all filters combined). We present the distributions relative to the two surveys in Fig.~\ref{fig:detections}.

Over the five-year window considered in this analysis, we predict to have 18000 SNe Ia with at least five LSST observations (in any filter) with $\mathrm{SNR}>3$. The majority of these SNe will have no detections from \euclid. This is expected given the sparse cadence of the \euclid\ mission. However, we predict that approximately 3700 SNe will have at least five \euclid\ detections with $\mathrm{SNR}>3$, and 1900 SNe will have at least ten detections with $\mathrm{SNR}>3$ from \euclid. 

In Fig.~\ref{fig:LC}, we show an example of a simulated light curve with LSST optical photometry and \euclid\ optical (VIS) and NIR (\emph{YJH}) photometry. We expect a few hundred SNe to have similar data coverage and quality in the NIR (as shown in Fig.~\ref{fig:detections}). In the next section, we show how the additional \euclid\ NIR data can significantly improve SN distance measurements inferred using LSST optical data alone.

\subsection{Supernova light-curve fitting and distances}
\label{sec:SNfit}
SNe Ia can serve as standard candles for measuring cosmological distances and constrain the expansion history of the Universe. Generally, we measure SN distances using the SN rest-frame B-band peak brightness.
However, not all SNe are always measured at maximum light and in their rest-frame B-band. For this reason, it is necessary to use SN Ia SED time-series models to perform light-curve fitting and determine SN light-curve properties and peak brightness.

For our analysis, we fit the simulated light curves using the same SALT3 model introduced in Sec~\ref{sec:sim_general}. In the fit, we assume the SN spectroscopic redshift is known and we only fit for the observed SN peak brightness in rest-frame B-band $m_B$ (also defined as $-$2.5 $\log_{10}(x_0)$), the SN light-curve stretch $x_1$, SN color $c$ and time of peak brightness $t_0$, and we estimate the relative uncertainties. 

\begin{figure}
    \centering
    \includegraphics[width=0.9\columnwidth]{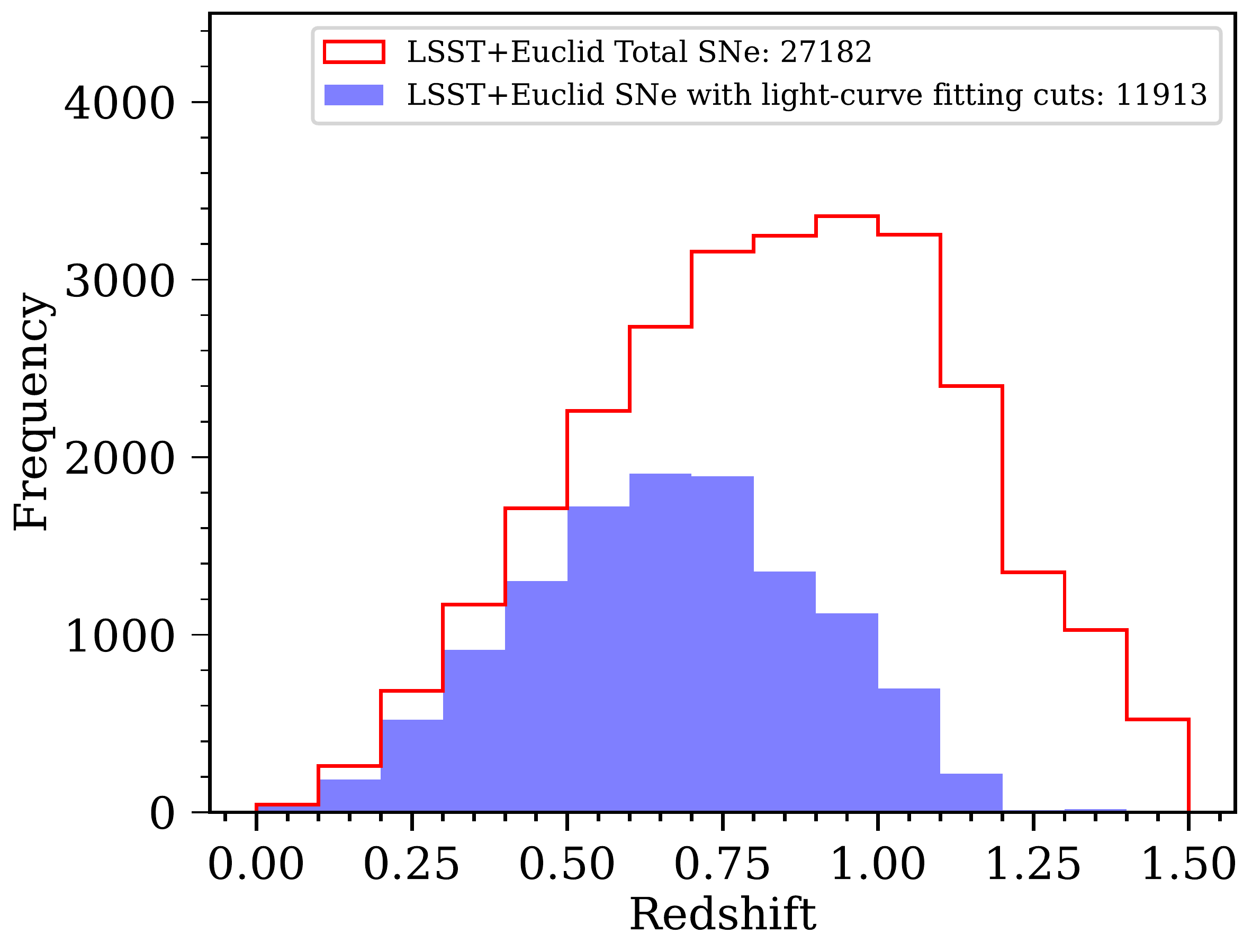}\\
    \caption{Redshift distribution of the SNe Ia successfully fitted with the SALT3 model, and SNe Ia passing the SALT-based quality cuts discussed in Sec.~\ref{sec:SNfit}. }
    \label{fig:RD}
\end{figure}

Using LSST and \euclid\ mock data combined, we find that approximately 27000 SNe Ia are successfully fitted with the SALT3 model, and 11000 of these SNe pass the light-curve fitting quality cuts generally applied in SN Ia cosmological analyses \citep[i.e., $-0.3<c<0.3$, $-3<x_1<3$, $\sigma_{x_1}<1$, and $\sigma_{t_0}<2$,][]{2014A&A...568A..22B}. In Fig.~\ref{fig:RD}, we present the redshift distributions of the fitted SNe Ia. 

Given the results from the light-curve fitting, we can standardize SN brightnesses and infer distances using the Tripp formula \citep{Tripp98, Astier06}:

\begin{equation}
\mu_{\mathrm{obs}} = m_B +\alpha x_1 - \beta c - \mathcal{M}_B,
\label{eq:mu}
\end{equation}
where $\alpha$ and $\beta$ are the color and stretch corrections coefficients and $\mathcal{M}_B$ is the SN Ia intrinsic brightness in rest-frame B-band.

\begin{figure*}
    \centering
   \includegraphics[width=0.8\textwidth]{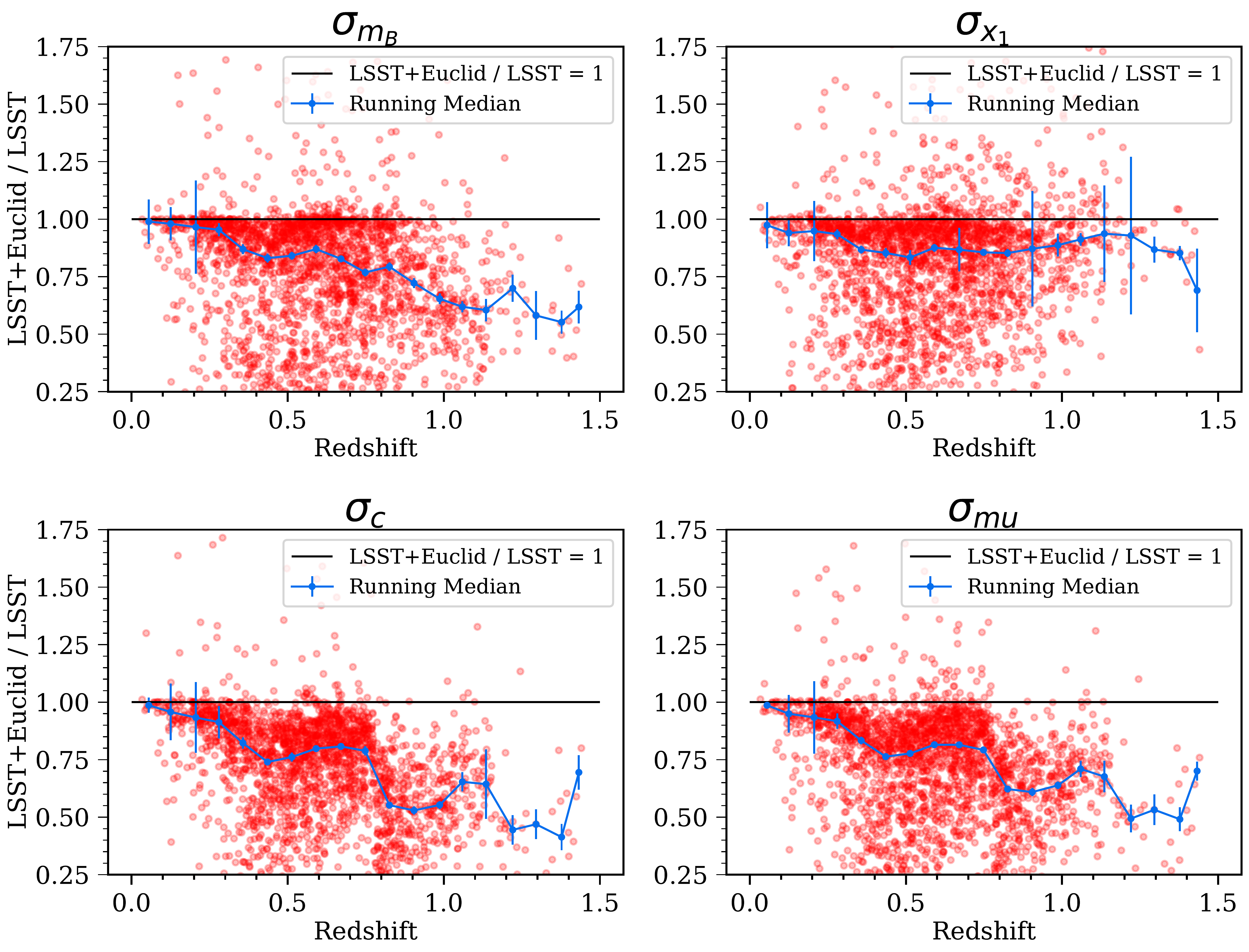}
    \caption{Fractional comparison of uncertainties on $m_B$ (upper left), $x_1$ (upper right), $c$ (lower left), $\mu$ (lower right) when including or not \euclid\ mock data. We present the fractional plots with the error in \euclid\ combined with LSST as a fraction of the error in LSST with a running median (blue points and line) of the binned data. We have also identified where the errors are equivalent and the fractional values, subsequently, equal to 1 (black horizontal line).}
    \label{fig:ERRcomp}
\end{figure*}

In Fig.~\ref{fig:ERRcomp}, we consider SNe with at least five \euclid\ observations with SNR$>3$ and we compare uncertainties on SN fitted parameters when considering LSST+\euclid\ and when considering LSST alone. We also estimate uncertainties on SN distances by propagating uncertainties on $m_B$, $x_1$ and $c$ and applying eq.~\ref{eq:mu}, and we compare results with and without \euclid\ mock data (see Fig.~\ref{fig:ERRcomp}, lower right panel).

We find that uncertainties on SN fitted parameters and SN distances are reduced compared to when including \euclid\ data compared to when using LSST data alone. The impact of the additional measurements from \euclid\ is strongest at higher redshift SNe ($z>0.8$), where the rest-frame optical SN flux starts to redshift into observer-frame Y and J bands, and SN flux in observer-frame $g$-band drops significantly, thus making fits of LSST-only light curves more uncertain (see Fig.~\ref{fig:bands_specia}). 
For $z>0.8$, we expect the additional \euclid\ mock data to reduce uncertainties on SN distances by 40--50\%.

\section{Constraining the SNe Ia \lq mass step\rq\ using LSST and \euclid\ mock data}\label{sec:Mass_Step}

Various SN Ia analyses have shown that the intrinsic brightness of SNe Ia correlates with host galaxy properties, and in particular SNe Ia found in more massive galaxies are more luminous than SNe Ia found in lower mass galaxies. The astrophysical origin of this correlation is still uncertain, therefore many SN Ia cosmological analyses have empirically modeled this dependency as a step function at host galaxy stellar mass $10^{10} M_{\odot}$. Analyzing a compilation of a thousand of spectroscopically confirmed SNe Ia with high quality optical data, \citet{BS20} showed how dust, and in particular different properties of dust for SNe in high and low-mass galaxies, can explain the observed mass step and other correlations between SNe Ia intrinsic scatter and SN color.

Rest-frame NIR SN data can provide compelling evidence for whether dust is the cause of the mass step. In fact, if dust is the main cause of the luminosity step observed in the optical, we expect the luminosity step in the NIR to be significantly reduced. \citet{Uddin20} and \citet{2021ApJ...923..197P} measured the mass step from SNe Ia with rest-frame NIR data. Both analyses measure mass steps larger than 0.07 in the rest-frame NIR, with $>2\sigma$ significance. These results were obtained analysing limited ($<$150 likely SNe Ia) low-$z$ SN samples with ground-based NIR data from the Carnegie SN project and other literature compilations \citep{2010AJ....139..519C, 2011AJ....142..156S, 2014ApJ...784..105W}.

As discussed in Sec.~\ref{sec:SNquality}, we expect the \euclid\ mission to obtain hundreds of high-quality SN light curves with good rest-frame NIR coverage and a well-understood selection function. In this section, we show how \euclid\ mock data, combined with optical data from LSST, can be used to give a definitive answer on the origins of the mass step.

\subsection{Simulating the SN Ia mass step}
For this part of our analysis, we generate three sets of LSST+\euclid\ simulations. We follow the same simulation framework presented in Sec.~\ref{sec:sim_general}, but we vary the modeling of SNe Ia intrinsic properties and the modeling of the mass step. The first simulation is generated as described in Sec.~\ref{sec:sim_general}, i.e. assuming the SN Ia intrinsic scatter model by \citet{Guy10}, assuming luminosity-stretch and luminosity-color correlations have slope $\alpha=0.14$ and $\beta=3.0$ and fixing the intrinsic mass step to zero. In the second simulation, SNe Ia are also generated following the intrinsic scatter model by \citet{Guy10} and including stretch and color correlations, but additionally we introduce a wavelength-independent (`grey') mass step of 0.08 mag at $10^{10} M_{\odot}$. 

In the third simulation, we include luminosity-stretch correlations ($\alpha=0.14$) and correlations between SN stretch and SN host mass, but we follow the modeling presented by \citet{BS20} to generate SN intrinsic colors  ($c_{\mathrm{int}}=-$0.084, $\sigma_{c}=-$0.042), intrinsic luminosity-color correlations ($\beta_{\mathrm{int}}=2.0$), extrinsic dust reddening \citep[$\tau_E=0.14$, see eq.~12 from][]{BS20} and dust total-to-selective absorption ($R_V=1.5$ in high mass galaxies, $R_V=2.75$ in low mass galaxies, with $\sigma_{R_V}=1.3$).
For each sample of simulated SNe Ia, we perform light-curve fitting and standardization in the rest-frame optical only and rest-frame NIR only (see Sec.~\ref{sec:opt_NIR_HD}) and measure the recovered mass step (see Sec.~\ref{sec:mass_step_meas}). 

\subsection{The optical rest-frame and NIR rest-frame Hubble diagrams}
\label{sec:opt_NIR_HD}
We fit simulated LSST and \euclid\ light curves using rest-frame optical data only ($3000-7000$ \AA) and NIR rest-frame data only ($1-2\mu m$). As shown in Fig.~\ref{fig:bands_specia}, with the \euclid\ mock data we can obtain good coverage in the rest-frame NIR up to $z=0.8$. 
We perform light-curve fitting using the SALT3 model by \citet{Kenworthy21} and \citet{Pierel22} (see also Sec.~\ref{sec:sim_general}). When using optical rest-frame data only, we fit for the four light-curve fitting parameters $m_B$ (or $x_0$), $x_1$, $c$ and $t_0$ (similarly to the approach presented in Sec.~\ref{sec:SNfit}). When using NIR rest-frame data only, we fit for the amplitude term $x_0$ only. The times of peak $t_0$ and the stretch values $x_1$ are \textit{fixed} to the values fitted from the rest-frame optical data, while SN color is fixed to zero. The higher cadence optical data provides significantly better constraints on time of peak compared to the lower-cadence NIR data, and NIR light curves are significantly less sensitive to stretch. 

After performing light-curve fitting, we select SNe that pass the SALT-based cuts discussed in Sec.~\ref{sec:SNfit}. We also apply a redshift cut at $z<0.8$ to reduce effects of selection biases.
In Table~\ref{tab:mass-step}, we present the numbers of SNe Ia for each generated simulation, both for optical rest-frame and NIR rest-frame fits. 

\begin{table}
    \centering
    \begin{tabular}{l|cccc}
                & $N_{\mathrm{SNe}}$ & $N_{\mathrm{SNe}}$ &          &                \\
 Simulation     & optical fit & NIR fit & $\gamma_{\mathrm{Opt}}$ & $\gamma_{\mathrm{NIR}}$ \\
 \hline

`zero' mass step & 8449 & 3932 & -0.001(3) & 0.006(13) \\
`grey' mass step & 8485 & 3965 & -0.081(3) & -0.075(11) \\
Dust-based mass step & 8036 & 3878 & -0.078(3) & -0.027(11) \\

\hline
    \end{tabular}
    \caption{Number of fitted SNe and recovered mass steps for each simulation tested.}
    \label{tab:mass-step}
\end{table}

Given the SN light-curve fitting, SNe Ia standardized distances are measured applying eq.~\ref{eq:mu}.
Modern cosmological analyses include in the estimation of $\mu_{\mathrm{obs}}$ corrections for selection effects (so-called `bias corrections') and determine the nuisance parameters $\alpha$, $\beta$ and $\mathcal{M}_B$ while performing cosmological fitting.
For simplicity, in our analysis we assume bias corrections to be negligible and we fix the nuisance parameters to the simulated values ($\alpha=0.14$, $\beta=3.0$ and $\mathcal{M}_B=-19.365$).

We define uncertainties on $\mu_{\mathrm{obs}}$ as
\begin{equation}
\begin{split}
\sigma_{\mu_{\mathrm{obs}}}^2= & \sigma_{m_B}^2 +(\alpha \sigma_{x_1})^2 + (\beta \sigma_{c})^2 \\ 
& +\alpha C_{x_1,mB} -\beta C_{c,mB} - \alpha\beta C_{x_1,c} +\sigma_{\mathrm{int}}^2,
\end{split}
\label{eq:muerr}
\end{equation}
where $\sigma_{\mathrm{int}}$ is the SN Ia intrinsic scatter which is fixed to the simulated value of 0.11, while $ C_{x_1,mB}$, $ C_{c,mB}$ and $ C_{x_1,c}$ are the fitted covariance matrices among the SALT3 parameters. This definition provides a realistic estimate of the expected SN distance uncertainties.

The SN Ia standardized distances and distance uncertainties are then used to build the redshift-distance diagram, usually referred to as a \lq Hubble diagram\rq , and constrain cosmological parameters.
The residuals between $\mu_{\mathrm{obs}}$ and distances predicted by the best-fit cosmology are usually defined as \lq Hubble residuals\rq . In our analysis, we do not perform the full cosmological fit of our simulated data and we simply define Hubble residuals as:
\begin{equation}
    \mu_{\mathrm{res}} = \mu_{\mathrm{obs}}-\mu_{\Lambda \mathrm{CDM}}
\end{equation}
where $\mu_{\Lambda \mathrm{CDM}}$ are SN distances predicted by the input cosmological model (rather than the best-fit one) used in our simulations.

In Fig.~\ref{fig:mass_step_dispersion}, we present simulated Hubble residuals as a function of SN host stellar mass. The typical dispersion of simulated Hubble residuals when using optical rest-frame data only is $\sim$0.16, while for NIR rest-frame data only the dispersion is larger ($>0.5$ mag). This affects our ability to constrain the mass step.

\subsection{Measuring the mass step}
\label{sec:mass_step_meas}
\begin{figure}
    \centering
    \includegraphics[width=0.5\textwidth]{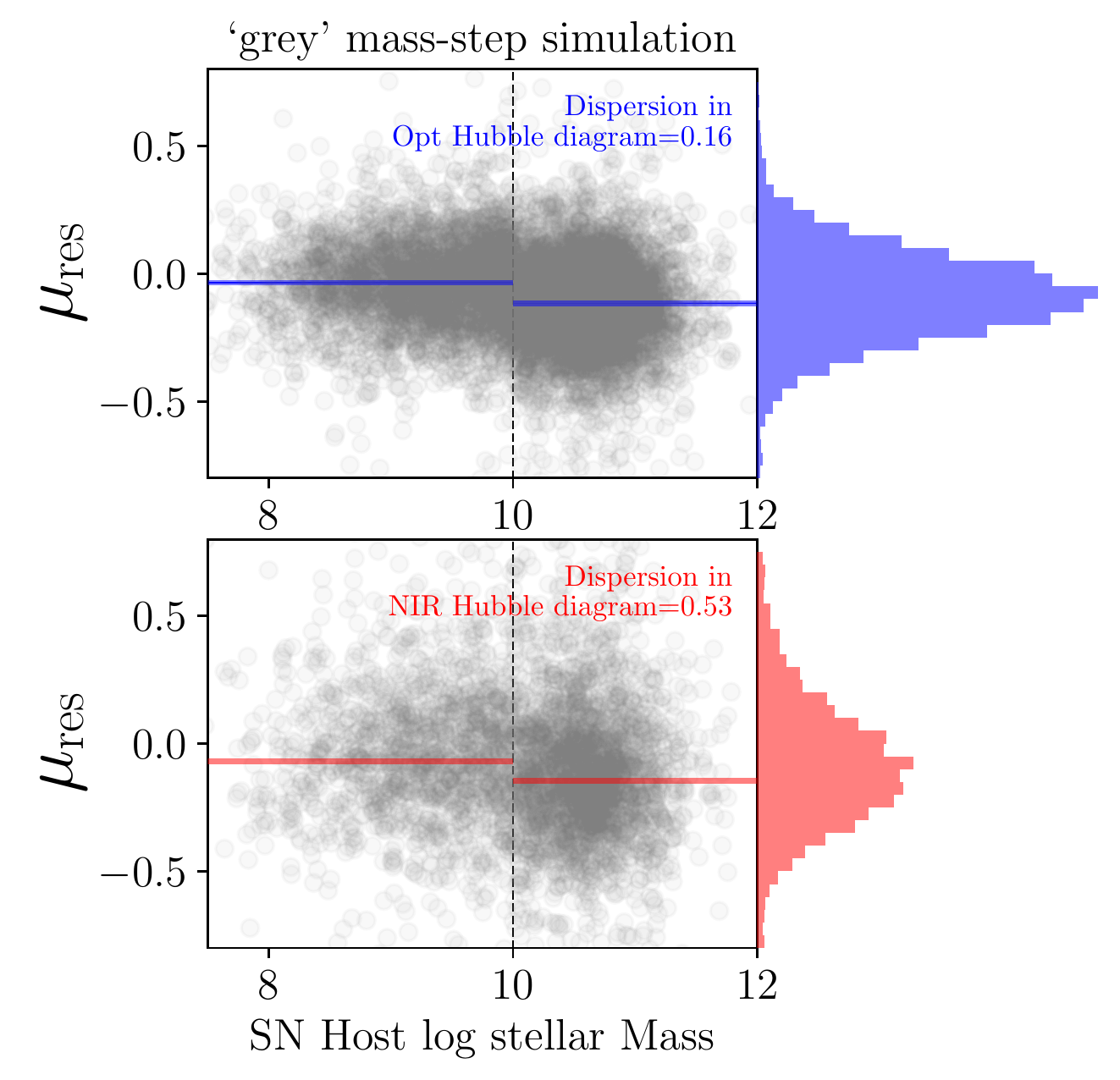}
    \caption{SN Hubble residuals as a function of SN host stellar mass, measured using rest-frame optical data only ($3000-7000$ \AA, upper plot) and rest-frame NIR data only ($1-2$ $\mu m$, lower plot). Dispersion on the Hubble diagram built using NIR rest-frame data only is three times larger compared to the Hubble diagram built using optical data. Uncertainties on the recovered mass steps are 0.003 and 0.011 from the optical and NIR Hubble residuals, respectively.}
    \label{fig:mass_step_dispersion}
\end{figure}

Given SN distances and their uncertainties, we measure the mass step as
\begin{equation}
    \gamma = \langle \mu_{\mathrm{res}} \rangle_{M_{*}>10^{10}M_{\odot}} - \langle \mu_{\mathrm{res}} \rangle_{M_{*}<10^{10}M_{\odot}} ,
    \label{eq:gamma}
    \end{equation}
where $\langle \mu_{\mathrm{res}} \rangle$ is the \textit{weighted} average of Hubble residuals, with weights defined as $1/\sigma_{\mu_{\mathrm{obs}}}^2$ (see eq.~\ref{eq:muerr}). In Fig.~\ref{fig:mass_step_dispersion}, we compare the mass step estimates for one of our simulations (`grey' mass step simulation) with the Hubble diagram dispersion. Despite the large dispersion in the NIR Hubble diagram, the large number of SNe Ia allows us to recover a mass step with a $\sim$0.01 mag uncertainty.

When considering the rest-frame optical SN fits, all the terms in equations \ref{eq:mu} and \ref{eq:muerr} are non-zero. For the rest-frame NIR SN fits, color corrections are equal to zero by definition (as SN $c$ is fixed to zero in the fits), as well as the stretch and color uncertainties $\sigma_{x_1}$ and $\sigma_{c}$ (as both $x_1$ and $c$ are not floated in the fit).

\begin{figure}
    \centering
    \includegraphics[width=0.43\textwidth]{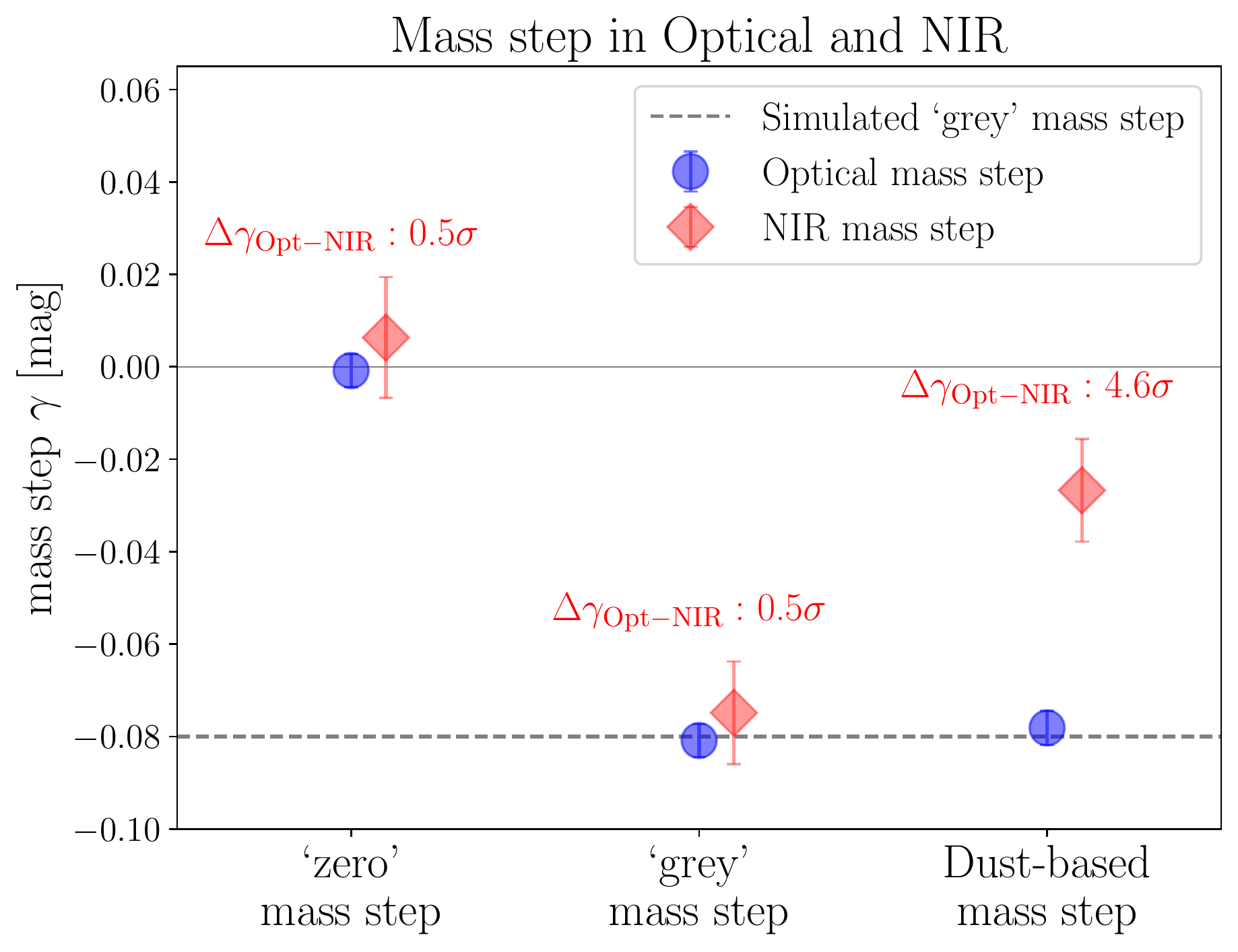}
    \caption{Mass step measured using rest-frame optical only ($3000-7000$~\AA, blue circles) and rest-frame NIR only ($1-2$ $\mu m$, red diamonds) data. We measure the recovered mass steps for three simulations: a simulation generated assuming the intrinsic scatter model presented by \citet{Guy10} with a zero mass step, a simulation generated assuming the intrinsic scatter model presented by \citet{Guy10} with a mass step of 0.05 mag, and a simulation generated assuming the dust-based scatter model presented by \citet{BS20}. 
    }
    \label{fig:mass_step}
\end{figure}

The recovered mass steps are presented in Fig.~\ref{fig:mass_step} and in Table~\ref{tab:mass-step}.
When considering optical rest-frame data, we find no significant mass step for the `zero' mass step simulation. For our `grey' mass step simulation and dust-based simulation, we recover a mass step of $-$0.081 $\pm$ 0.003 and $-$0.078 $\pm$ 0.003, in good agreement with the simulated value ($\gamma=0.08$).
When considering NIR rest-frame fits, we again find no significant mass step for the `zero' mass step simulation. For our `grey' mass step simulation we have a mass step of $-$0.075 $\pm$ 0.011
, in excellent agreement with the simulated value and with the mass step recovered in the optical.
Finally, when considering our dust-based simulation, we find that the recovered mass step is $-$0.027 $\pm$ 0.011. 
This $4\sigma$ difference between the mass step recovered using optical-only and NIR-only rest-frame data constitutes one of the main results of this paper and it demonstrates that using Rubin and \euclid\ mock data we will be able to confirm (or rule out) with high confidence whether the (optical and NIR) mass steps are well-described by a dust-based model or requires an alternative astrophysical explanation.

\section{Discussion and Conclusions}\label{sec:Discussion}

We exploit public information from previous transient analyses within LSST and \euclid\ and produce simulations of SN Ia light curves as measured by the joint LSST and \euclid\ surveys. While the cadence is non-ideal for supernova studies, the \euclid\ Deep fields offer short windows where high-cadence, high SNR, measurements can be obtained. We study two utilities of the \euclid\ data. The first is combining \euclid\ NIR data with LSST optical data. As shown in Fig.~\ref{fig:ERRcomp}, this significantly improves the constraining power on distances at high redshifts, by as much as a factor of $2\times$ at $z \gtrsim 1$. The second utility is in a stand-alone \euclid\ measurement of the canonical `mass-step'. This NIR measurement has significant implications for one of the largest systematics for optical supernova studies.  We find that we can measure a predicted mass step by up to $4\sigma$ and can distinguish between different models (dust and non-dust driven) at $4\sigma$. This will be a first-of-its-kind measurement with high statistics at high redshift.

This analysis is complementary to those of \citet{Inserra18}, \citet{2022arXiv220408727M} and \citet{2022arXiv220409402T}; it uses only public information on the LSST and \euclid\ observing strategies and it is designed so that the community could quickly use this to forecast other transient studies, like what was done for the LSST Photometric LSST Astronomical Time Series Classification Challenge \citep[PLAsTiCC,][]{Kessler19}. As both LSST and \euclid\ are in the process of finalizing their observing strategies, our results are still preliminary. In Appendix \ref{appendix:snana_inputs}, we discuss our data release and our simulation inputs.

The assumption of this analysis is that SNe can be detected and measured with \euclid. While this absolutely can be done, the data must be processed in specific ways that are not applicable to static science surveys \citep[see][ for a description of what is planned for LSST]{Sanchez22}.  
We hope that this study will provide useful additional information to the \euclid\ and LSST consortia as they finalize their observing strategies and data-processing pipelines.

\section*{Acknowledgements}

The authors acknowledge Isobel Hook for the very useful discussions on the \euclid\ observing strategy and on this work. We also thank Rick Kessler for his support with the \snana\ simulation package. D.S. is supported by Department of Energy grant DE-SC0010007, the David and Lucile Packard Foundation, and the Sloan Foundation. This manuscript is based upon work supported by the National Aeronautics and Space Administration (NASA) under Contracts NNG16PJ34C and NNG17PX03C issued through the {\it Roman} Science Investigation Teams Program.  D.S.\ is supported in part by NASA grant 14-WPS14-0048. J.R.~was supported by NASA ROSES  12-EUCLID12- 0004 and some of this work was done at the JPL, which is run under a contract for NASA by Caltech.

%%%%%%%%%%%%%%%%%%%%%%%%%%%%%%%%%%%%%%%%%%%%%%%%%%
\section*{Data Availability}

We release our simulated catalogs, as well as all the \snana\ input files needed to reproduce our results in: \url{https://github.com/maria-vincenzi/Euclid_LSST_SNIaSims}.



\bibliographystyle{mnras}
\bibliography{main} % if your bibtex file is called example.bib

%\begin{thebibliography}{99}
%\end{thebibliography}

%%%%%%%%%%%%%%%%%%%%%%%%%%%%%%%%%%%%%%%%%%%%%%%%%%

%%%%%%%%%%%%%%%%% APPENDICES %%%%%%%%%%%%%%%%%%%%%

\appendix

\section{\snana\ inputs}
\label{appendix:snana_inputs}
All the \snana\ input files necessary to create the simulations presented in this paper are available on GitHub (\url{https://github.com/maria-vincenzi/Euclid_LSST_SNIaSims}).
Table~\ref{tab:SNANA_inputs} presents a summary of the files needed to reproduce the simulations and their usage.
\begin{table*}
    \centering
    \caption{Inputs of the SNANA simulations.}
    \begin{tabular}{|p{7.5cm}|p{6cm}|p{2cm}|}
        File Name & Usage & \snana\ key \\
        \hline
       - \texttt{EUCLID\_LSST\_DDF.INPUT}  & \textbf{Master file} used to generate all simulations & \texttt{Input file}\\
       - \texttt{SIMLIB\_baseline\_v2.0\_10yrs\_DDF.simlib.COADD}  &  LSST cadence &  \texttt{SIMLIB\_FILE}\\
       - \texttt{SIMLIB\_baseline\_v2.0\_10yrs\_DDF\_wEUCLID.simlib.COADD}  & LSST and \euclid\ cadences combined &  \texttt{SIMLIB\_FILE}\\
       - \texttt{kcor\_LSST\_EUCLIDvis.fits}  & LSST and \euclid\ filters and filter calibration &  \texttt{KCOR\_FILE}\\
       - \texttt{DES\_SVA2+LOGMASS\_LOGSFR\_Sullivan10.HOSTLIB}  & Host galaxy library for SN host simulation &  \texttt{HOSTLIB\_FILE}\\
       - \texttt{sn\_ia\_salt2\_g10.input}  & Simulate SN Ia intrinsic properties using intrinsic scatter model by \citet{Guy10}&  \texttt{INPUT\_FILE\_INCLUDE}\\
       - \texttt{sn\_ia\_salt2\_bs20.input}  & Simulate SN Ia intrinsic properties using intrinsic scatter model by \citet{BS20}& \texttt{INPUT\_FILE\_INCLUDE}\\
       - \texttt{DES\_WGTMAP\_MassSFR\_Wiseman2021.HOSTLIB}  & SN Ia rates as a function of host properties (no mass step) & \texttt{HOSTLIB\_WGTMAP\_FILE} \\
       - \texttt{DES\_WGTMAP\_MassSFR\_Wiseman2021\_STEP.HOSTLIB}  & SN Ia rates as a function of host properties and 0.08 mag mass step & \texttt{HOSTLIB\_WGTMAP\_FILE} \\
       - \texttt{SIMGEN\_TEMPLATE\_LSST\_EUCLID.INPUT}  & General set up & \texttt{SIMGEN\_INFILE\_Ia}\\
    \end{tabular}
    \label{tab:SNANA_inputs}
\end{table*}
%If you want to present additional material which would interrupt the flow of the main paper, it can be placed in an Appendix which appears after the list of references.

%%%%%%%%%%%%%%%%%%%%%%%%%%%%%%%%%%%%%%%%%%%%%%%%%%

% Don't change these lines
\bsp	% typesetting comment
\label{lastpage}
\end{document}